\documentclass[journal,letters]{IEEEtran}

\usepackage[final]{graphicx}
\graphicspath{ {./figs/} }
\usepackage[reqno]{amsmath}
\usepackage{amssymb}
\usepackage{amsthm}
\usepackage{subfig}
\usepackage{epstopdf}
\usepackage{algorithm}
\usepackage{algorithmicx}
\usepackage{algpseudocode}
\usepackage{setspace}
\usepackage[T1]{fontenc}
\usepackage{color,soul}
\usepackage{multirow}
\usepackage{array}
\usepackage{cite}
\usepackage{svg}
\usepackage{verbatim}

\usepackage{array}
\newcommand{\PreserveBackslash}[1]{\let\temp=\\#1\let\\=\temp}
\newcolumntype{C}[1]{>{\PreserveBackslash\centering}p{#1}}
\newcolumntype{R}[1]{>{\PreserveBackslash\raggedleft}p{#1}}
\newcolumntype{L}[1]{>{\PreserveBackslash\raggedright}p{#1}}

\begin{document}

\title{Knowledge Distillation For Wireless Edge Learning}

\author{ \large{Ahmed P. Mohamed, {\em Student Member, IEEE}}, \large{Abu Shafin Mohammad Mahdee Jameel, {\em Student Member, IEEE}}, \large{Aly El Gamal, {\em Senior Member, IEEE} }
\thanks{The authors are with the School of Electrical and Computer Engineering of Purdue University, West Lafayette, IN (e-mail: \{mohame23, amahdeej, elgamala\}@purdue.edu).
}}


\maketitle

\begin{abstract}
In this paper, we propose a framework for predicting frame errors in the collaborative spectrally congested wireless environments of the DARPA Spectrum Collaboration Challenge (SC2) via a recently collected dataset. We employ distributed deep edge learning that is shared among edge nodes and a central cloud. Using this close-to-practice dataset, we find that widely used federated learning approaches, specially those that are privacy preserving, are worse than local training for a wide range of settings. We hence utilize the synthetic minority oversampling technique to maintain privacy via avoiding the transfer of local data to the cloud, and utilize knowledge distillation with an aim to benefit from high cloud computing and storage capabilities. The proposed framework achieves overall better performance than both local and federated training approaches, while being robust against catastrophic failures as well as challenging channel conditions that result in high frame error rates.
\end{abstract}


\section{Introduction}

\IEEEPARstart{W}{ireless} edge devices are becoming ubiquitous for applications like  the  Internet  of  Things  (IoT) and smart cities. Due to the massive data generated by these edge devices in environments that are typically difficult to model, deep learning algorithms are ideal candidates for their intelligent  decision making to enhance the performance metrics for accuracy, safety and latency. Conventionally, effective training of deep learning models needs transferring the entire training dataset of all devices to a cloud server. Then, the cloud sends the trained model to all devices for on-device inference. However, with the huge increase in edge network sizes, this approach may not be suitable with the expanding requirements of real-time applications. Transferring training data to the cloud server may become unfeasible due to bandwidth limitation, may result in unexpected latency, and data privacy and security regulations may be violated. Hence, it is preferred to store and process data close to or at the devices. Consequently, a distributed training framework for sensitive data called federated learning was introduced, enabling training large numbers of edge models even with unbalanced and/or non independently and identically distributed (i.i.d.) data. 

The first design of federated learning is Federated Averaging (FedAvg) \cite{mcmahan2017communication}. In each communication round, the cloud sends the global model to a small set of randomly selected devices. Then, each of these devices updates the model based on its own training data and sends the trained model back to cloud, where new models are aggregated to obtain the updated global model. Finally, the cloud sends the aggregated model for all devices for inference. However, this approach suffers from privacy issues \cite{melis2019exploiting} as the model can leak information about the training data of the individual devices\cite{shokri2017membership}. Differential privacy based strategies were hence incorporated \cite{mcmahan2017learning}.

Obtaining better models is the primary stimulant for devices to participate in federated learning. However, federated models typically suffer from bias towards data distributions at selected devices, which can severely affect generalization \cite{agnostic_federated_learning}.

In this work, we utilize edge learning for frame error prediction using a DARPA Spectrum Collaboration Challenge (SC2) dataset. In this collaborative intelligent spectrum sharing environment, dynamic changes make this a difficult problem to solve using rule based or pure statistical modeling, and deep learning based methods were recently shown to deliver promising performance \cite{jameel2020deep}. The physical separation between communication nodes makes them subject to different local environments, which could aggravate the aforementioned federated learning generalization issue. 

Using the SC2 dataset, when measuring the classification accuracy of federated models for individual devices, we find that it is often significantly worse than that of locally trained models, eliminating the primary stimulant to take part in federated learning. Interestingly, we find that the inferior accuracy compared to local training is particularly significant in cases where channels exhibit a high frame error rate.

We hence propose a framework that depends on knowledge distillation to efficiently exploit the available high cloud computing and storage capabilities. To preserve privacy, this framework does not expose local data to the cloud, rather utilizing synthetically generated data to train the cloud teacher model. Our proposed framework achieves overall better performance than local and federated training alternatives. Compared to federated methods, it demonstrates robustness against catastrophic failures, where the distributed model is significantly outperformed by local training, and significantly outperforms federated models in highly erroneous environments.

\begin{figure}
    \centering
    \includegraphics[width=0.45\textwidth]{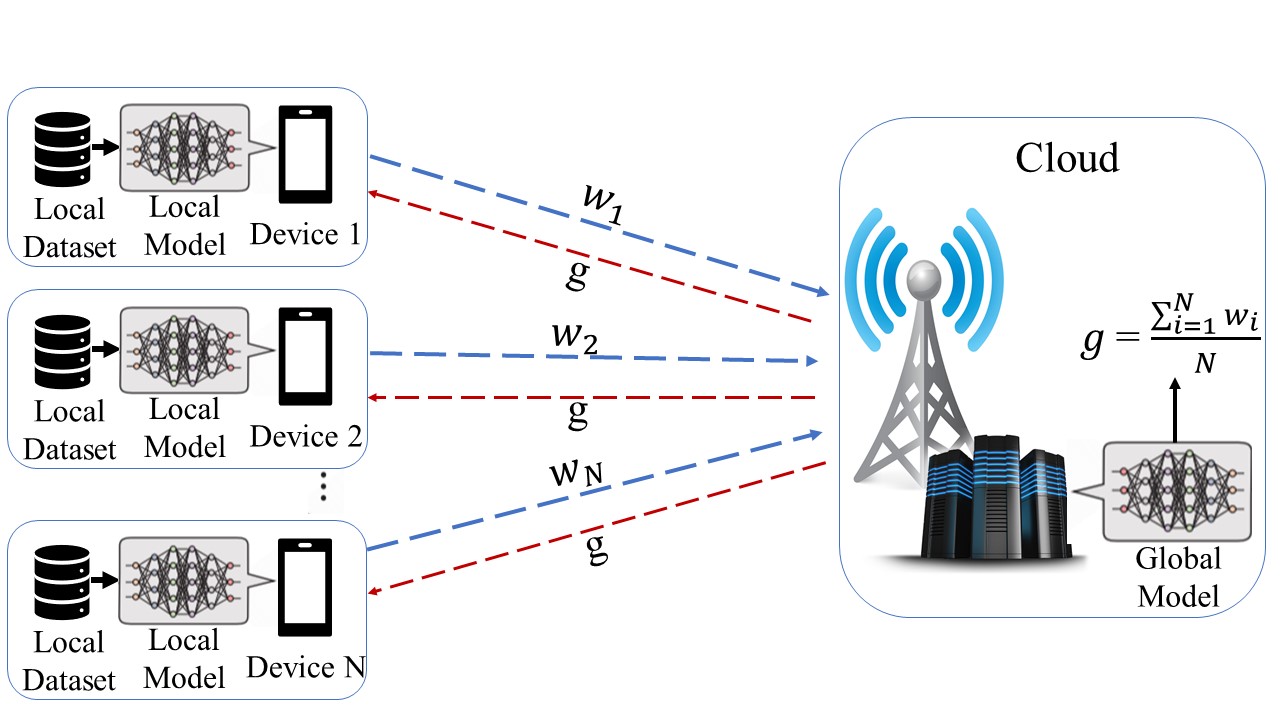}
    \caption{Federated Learning Model.}
    \label{fig:fd}
\end{figure}

\begin{figure}
    \centering
    \includegraphics[width=0.45\textwidth]{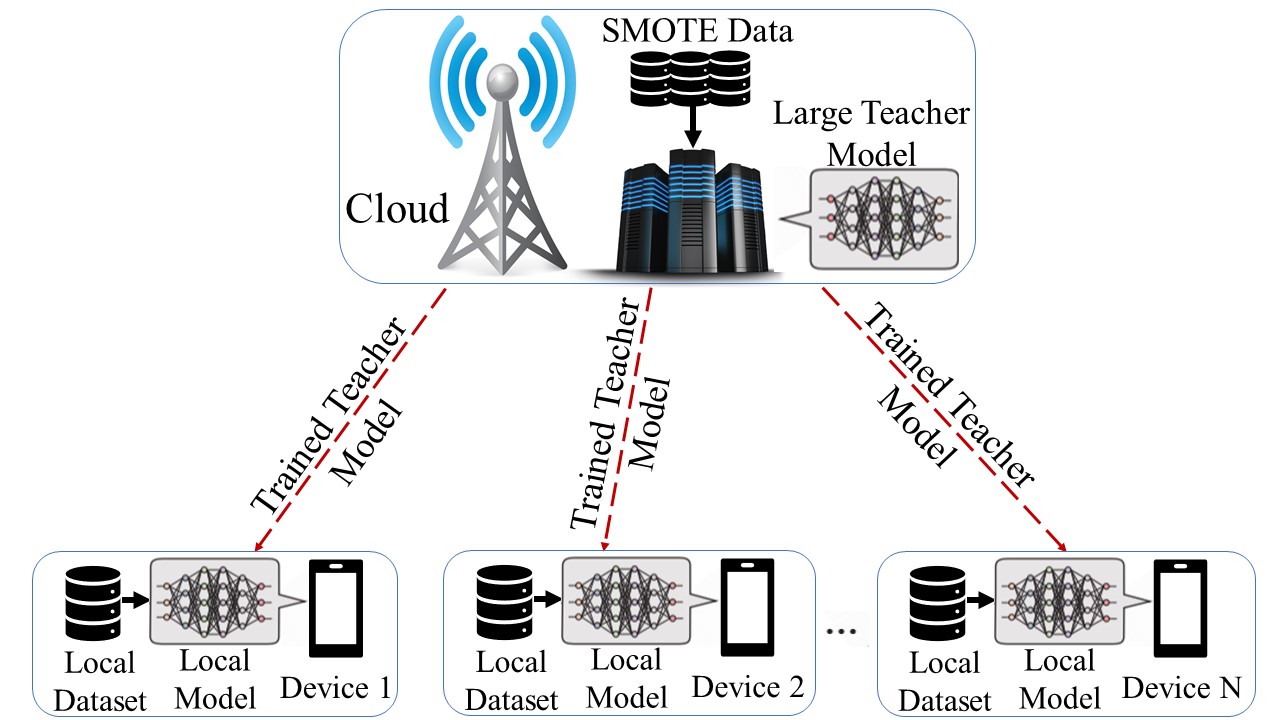}
    \caption{Proposed Knowledge Distillation Model.}
    \label{fig:kd}
\end{figure}
\section{Background}
\textbf{Federated Averaging (FedAvg):} is the original design for federated learning. It is a decentralized learning technique that utilizes a periodic model parameters exchange between the devices and the cloud without sharing information about the training data. First, the cloud initializes a global model $g_t$, randomly selects a small set of devices and sends them the current model. Then, each selected device with index $i$ uses its own training data to train this model as shown in Fig. \ref{fig:fd} and sends the learned parameters $w_i$ back to the cloud for aggregation into the new global model $g_{t+1}$:
\begin{equation}
g_{t+1} = \frac{\sum_{i=1}^{N} w_i}{N}. 
\end{equation}

\textbf{Differential Privacy Federated Averaging (DP-Fed):} is the original FedAvg technique incorporating differential privacy as a solution to solve the privacy issue. DP-Fed commonly uses Differentially
Private Stochastic Gradient Descent (DP-SGD) \cite{abadi2016deep} which is a noisy variant of the Stochastic Gradient Descent (SGD) optimizer. The key idea relies on clipping the gradients of the parameters then adding noise.    

\textbf{Synthetic Minority Oversampling Technique (SMOTE):} is a technique proposed to deal with class imbalance in a dataset by oversampling the minority class to make the dataset balanced. We utilized the idea of this technique to generate full synthetic training data for each device. First, SMOTE selects a random sample of data \textbf{\textit{v}}. Then, it selects $k$ (we selected $k=5$) neighbours for \textbf{\textit{v}} as described in \cite{chawla2002smote}. It randomly chooses one of the selected neighbours \textbf{\textit{w}} and generates a new sample \textbf{\textit{z}} as follows:
\begin{equation}
z = v + r * (v-w),
\end{equation}
where $r$ is a random number between 0 and 1.
Finally, each device sends its SMOTE generated training data to the cloud where it is stored and processed.

\textbf{Knowledge Distillation (KD):} was originally proposed as a compression technique to distill knowledge from a huge trained teacher model to a small student model without the need for costly computation and storage\cite{hinton2015distilling}. The basic idea for KD is that the student model not only trains through information provided by class labels, but also with the help of softened probabilities provided by the teacher model output, as depicted in Fig. \ref{fig:kd}. Let the logits (inputs to the final softmax layer) for both the teacher and student models be $o_t$ and $o_s$, respectively. For classical student model training, the cross entropy loss $l_{cross}$ is used to penalize the mismatch between the student model output $\sigma(o_s)$ and the class label $y_s$. In other words, the loss of the student model in this case can be represented as
\begin{equation}
l_S = l_{cross}(\sigma(o_s),y_s).
\end{equation}
In KD, the student model attempts to match its softened output $\sigma(o_s/T)$ with the softened output of the teacher model $\sigma(o_t/T)$ using a KL-Divergence loss,
\begin{equation}
l_{KL} = T^2 KL(\sigma(o_s/T),\sigma(o_t/T)),
\end{equation}
where T is a hyperparameter known as a temperature, which was originally introduced to provide control over the softening probability distribution of the teacher model output. The student model in KD is trained using the following loss function,
\begin{equation}
l_{KD} = \alpha l_S + (1-\alpha)l_{KL},
\end{equation}
where $\alpha$ is a weight hyperparameter.

\textbf{Transfer Learning with KD (TF-KD):} We develop a technique integrating transfer learning with KD. First, each device trains using KD with only its own generated SMOTE data. Then, each device fine-tunes its trained KD model by training it locally on its original local data. 

\section{Dataset}

\begin{figure}
    \centering
    \includegraphics[width=0.5\textwidth]{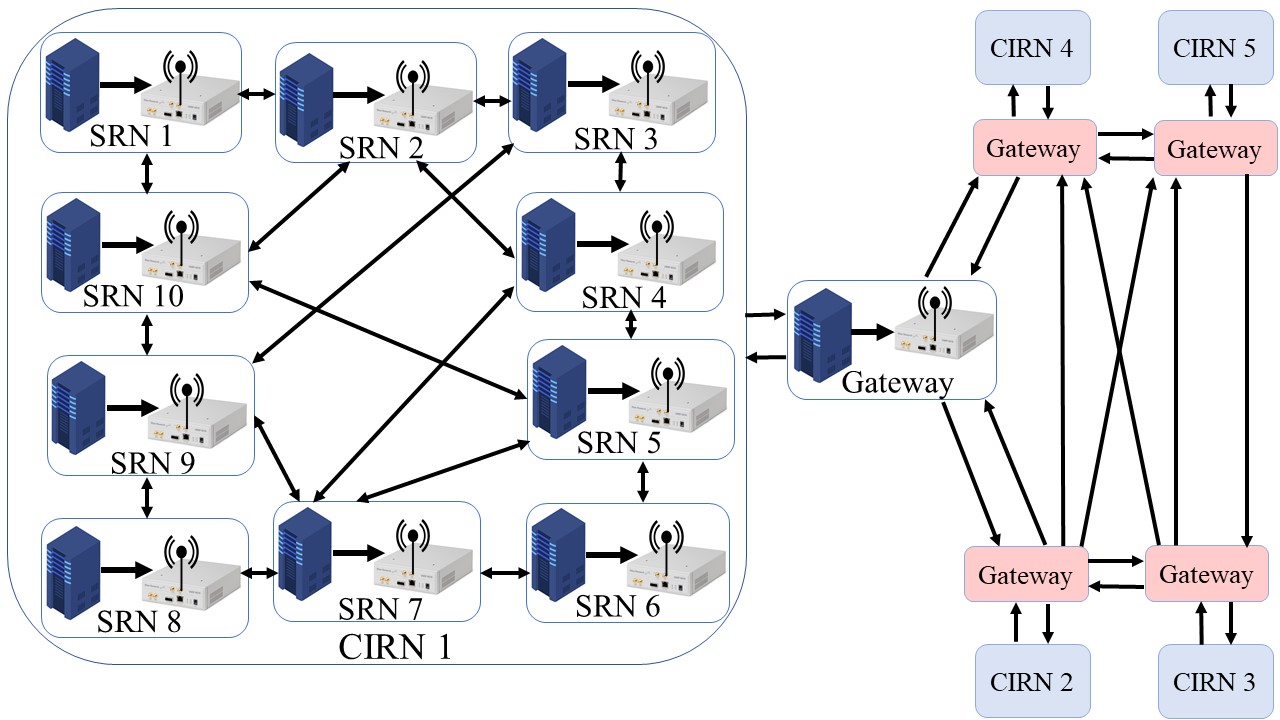}
    \caption{Match setup of DARPA SC2 Dataset involving 5 CIRNs with 10 nodes each, with details shown for CIRN 1.}
    \label{fig:sysbd}
\end{figure}

In this work, we utilize a version of our recently collected SC2 dataset used in \cite{jameel2020deep}\footnote{The dataset is collected from the third and final phase of DARPA SC2 and is available publicly with documentation as well as the technical design report for our SC2 Collaborative Intelligent Radio Network (CIRN) at https://github.com/amahdeej/sc2-frame-error. The source code for this work is also available at https://github.com/amahdeej/sc2-edge-learning}.
The dataset contains telemetry from wireless nodes in a CIRN deployed in a specific environment (scenario) with other competing Collaborative Intelligent Radio Network (CIRN)s. The gateways of the CIRNs use a common protocol, called the CIRN Interaction Language (CIL) to coordinate the dynamic allocation of channels (see Fig. \ref{fig:sysbd}). A CIRN is designed to extract maximum QoS performance, while ensuring that other CIRNs have resources available to maintain a baseline QoS. There are 14 scenarios in total, ranging from urban, rural and high path loss scenarios to specialized environments. For each match, a CIRN consists of 10 Software Radio Nodes (SRN), and 5 CIRNs share the spectrum in a band-congested environment. The dataset is divided into two parts, Scrimmage 4 (Scr4) and Scrimmage 5 (Scr5), which correspond to the last two practice rounds (Scrimmages) before the final event of SC2. In Scr4, the channel allocation decision parameters for our CIRN are randomized\footnote{See \cite{bamTech} for an explanation of bandwidth and channel allocation factors.}, leading to more frequent re-allocation. We accumulate data for different nodes (edges) across matches, resulting in 154 nodes for Scr4 and 408 nodes for Scr5. The randomization strategy in Scr4 results in a higher frame error rate. The frame error rates for Scr4 and Scr5 are 39\% and 30\%, respectively, resulting in a combined frame error rate of 33.33\%.

\section{Edge Learning for Frame Error Prediction}

In this collaborative spectrum sharing communication environment, the dynamically changing environmental parameters make it hard for traditional frame error prediction methods to perform well, and deep learning based methods offer promising performance improvements. Frame error prediction is vital for individual nodes to obtain the best possible QoS, and autonomously select the best possible transmission parameters. This frame error prediction problem is approached at the transmission side, using data that is available only to the transmitting node. If the current parameters indicate a possibility of frame error at the receiver side, the transmitting node can then tune the transmission parameters to avoid the error, saving vital resources like channel bandwidth and processing at the receiver side. We utilize the frame transmission error information available from the receiver side in the dataset to train and test our prediction algorithms.

Our node based dataset is ideal for edge learning. Unlike synthetic benchmarks where the data is arbitrarily divided into groups for edges, the dataset is built from the ground up using per edge data collection. Each node is subject to widely different channel parameters determined by DARPA's realistic Colosseum emulator, thereby creating variability in local data that can impact decisions. On the other hand, aggregation of information from nodes allows us to train using a much diverse information set. In a communication network like this, ensuring data privacy as well as maintaining generalization performance when aggregating information are also major issues. 
The goal of this study is to offer a framework that can utilize aggregated information in an edge learning environment to improve frame error prediction accuracy compared to local training while preserving privacy. 

To predict frame errors, we utilize the noise variance, modulation and coding scheme and allocated channel information, where these features were selected based on a Recursive Feature Elimination (RFE) saliency analysis. Keeping the feature vector compact minimizes the required computational cost. In our tests, this compact feature vector demonstrated performance that is close to larger feature vectors containing further information such as the Power Spectral Density (PSD) measurements. The test train split is kept as 1:1, i.e. half the data from each edge or node is used for training. Due to the inherent differences between Scr4 and Scr5, all models are trained and tested separately on them.

The network used at individual edges for local training is based on a multi layer perceptron (MLP) based approach. The considered MLP is similar to the one in \cite{jameel2020deep} and has a small size\footnote{MLP and CBSDNN have $7952$ and $247950$ trainable parameters, in order.}, hence its training is computationally cheap, which makes it suitable for use at nodes with low processing power. We used the same network for the considered federated approaches. Further, for federated approaches, 10 communication rounds are employed. For DP-Fed, We use Gaussian noise $N(0,\sigma$) with standard deviation $\sigma = 0.01$. For knowledge distillation based approaches, the student networks are MLP based. The teacher network is a CBSDNN network similar to the one used in \cite{jameel2020deep}, which combines the advantages of CNN layers with bidirectional Simple Recurrent Units (SRU). 
Based on empirical testing, we use $T = 10$, $\alpha = 0.4$ for Scr4 and $\alpha = 0.5$ for Scr5. For MLP model training, we use the SGD optimizer with a learning rate of $.001$ and a momentum of $0.9$ and for CBSDNN, we use the Adam optimizer with a learning rate of $.0001$. 

The list of abbreviations for considered approaches is as follows. \textbf{Local}: Local Training Only.
\textbf{FedAvg}, \textbf{DP-Fed} and \textbf{TF-KD} are as defined above. \textbf{KD-Scr}: Knowledge Distillation using real data for teacher network training. \textbf{KD-SMOTE}: Knowledge Distillation using SMOTE data for teacher network training. \textbf{Ensemble}: Ensemble method taking votes from three privacy preserving methods (Local, DP-Fed, KD-SMOTE).

\section{Results}

\begin{figure}
    \centering
    \includegraphics[width=0.45\textwidth]{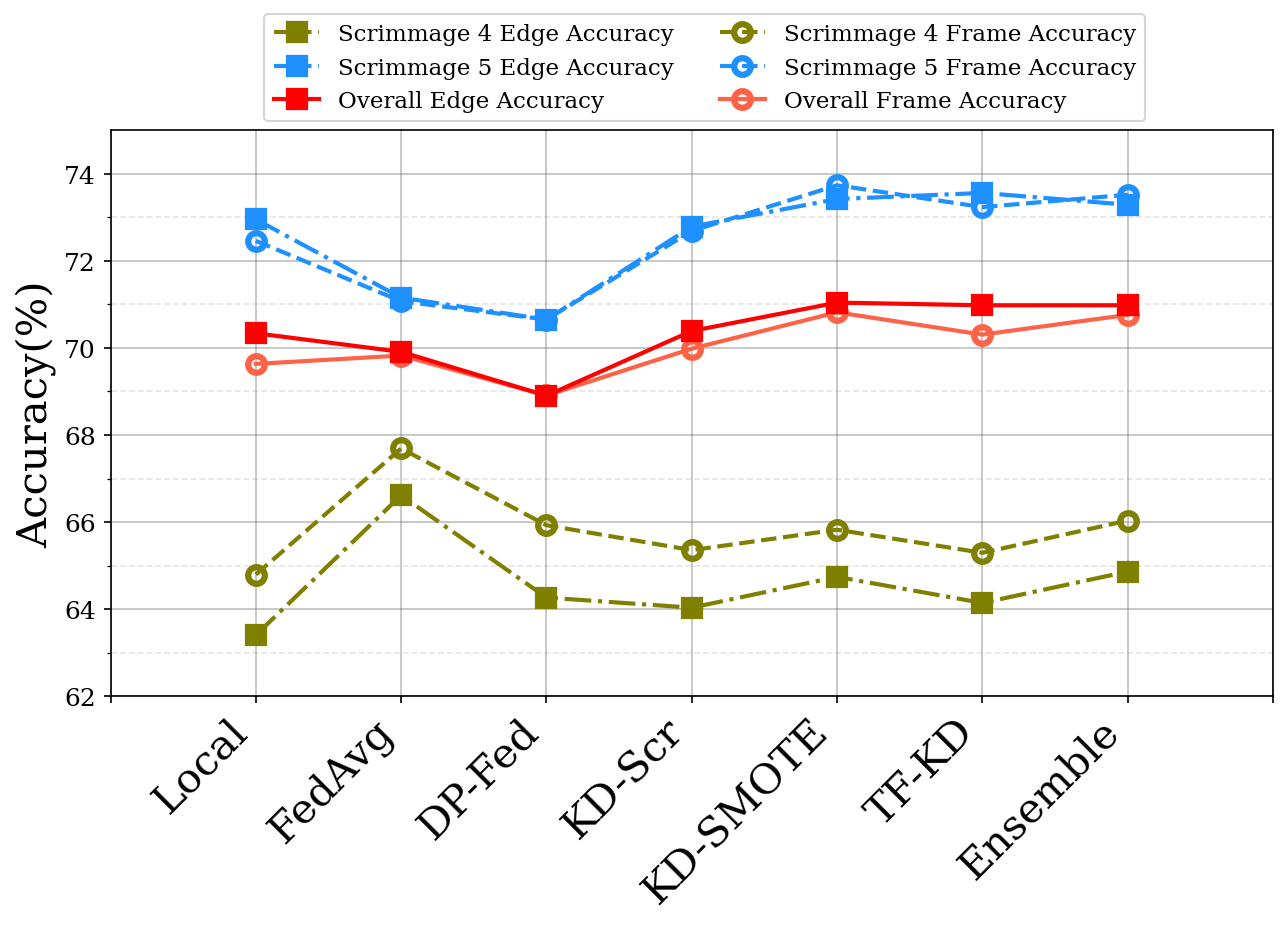}
    \caption{Accuracy for different algorithms.}
    \label{fig:Accr}
\end{figure}

In Fig. \ref{fig:Accr}, the accuracy of considered algorithms is presented for scrimmages 4 and 5. Here, Edge Accuracy represents the average of individual edge accuracy, while Frame Accuracy represents the average accuracy of all frames across edges. In Scr4, the factors behind channel allocation are randomized based on either the PSD measurement or the received CIL message, thereby exploring more unique channels per match compared to Scr5. This manifests itself in a consistently higher frame error rate, as randomization typically requires more frames for effective training. When the overall accuracy of both Scrimmages 4 and 5 is considered, KD based methods outperform other methods. On an overall frame accuracy metric, federated learning outperforms local training, but on the average edge accuracy metric, it falls below. The privacy-preserving DP-Fed results in a further drop. On the other hand, moving from KD-Scr to privacy preserving KD-SMOTE and TF-KD confers minimal performance change, and outperforms local accuracy over both edge and frame metrics.

\begin{figure}
    \centering
    \includegraphics[width=0.45\textwidth]{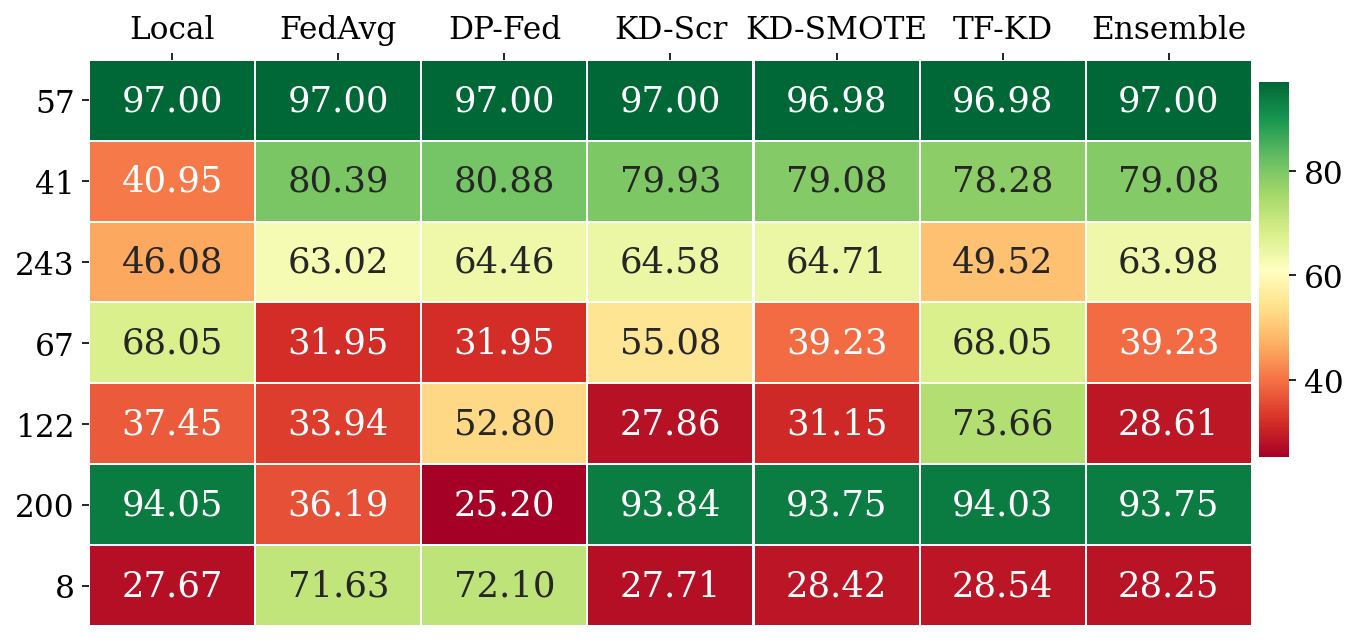}
    \caption{Accuracy for representative example nodes in Scrimmage 5.}
    \label{fig:Nodes}
\end{figure}

It is interesting to examine results for individual nodes - depicted in Fig. \ref{fig:Nodes} - that are representative of interesting cases as follows. Node 57 in the figure represents a case where all considered algorithms deliver acceptable performance. For Node 41, all distributed methods outperform local training. For Node 243, the TF-KD method fails along with local training. On the other hand, for Node 67, TF-KD and local training deliver significantly superior performance than all other distributed methods. For Node 122, only TF-KD delivers good performance. For Node 200, we can see a case where federated methods fail while others do not. Finally, for Node 8, federated methods are the only successful ones.

In summary, \textbf{the complementarity in performance between KD and federated based methods} is interesting and could pave the way for novel ensemble approaches that capitalize on this observation. In particular, it is obvious how KD based methods, and specially TF-KD, succeed in cases where local training outperforms federated learning. Furthermore, by analyzing the cases represented through nodes 67, 122 and 200, we can see that as \textbf{TF-KD combines local and distributed training, it can succeed when either of them fails}. This is interesting, specially as the considered ensemble method also combines local and distributed training but fails in most of these cases.

\begin{figure}
    \centering
    \includegraphics[width=0.45\textwidth]{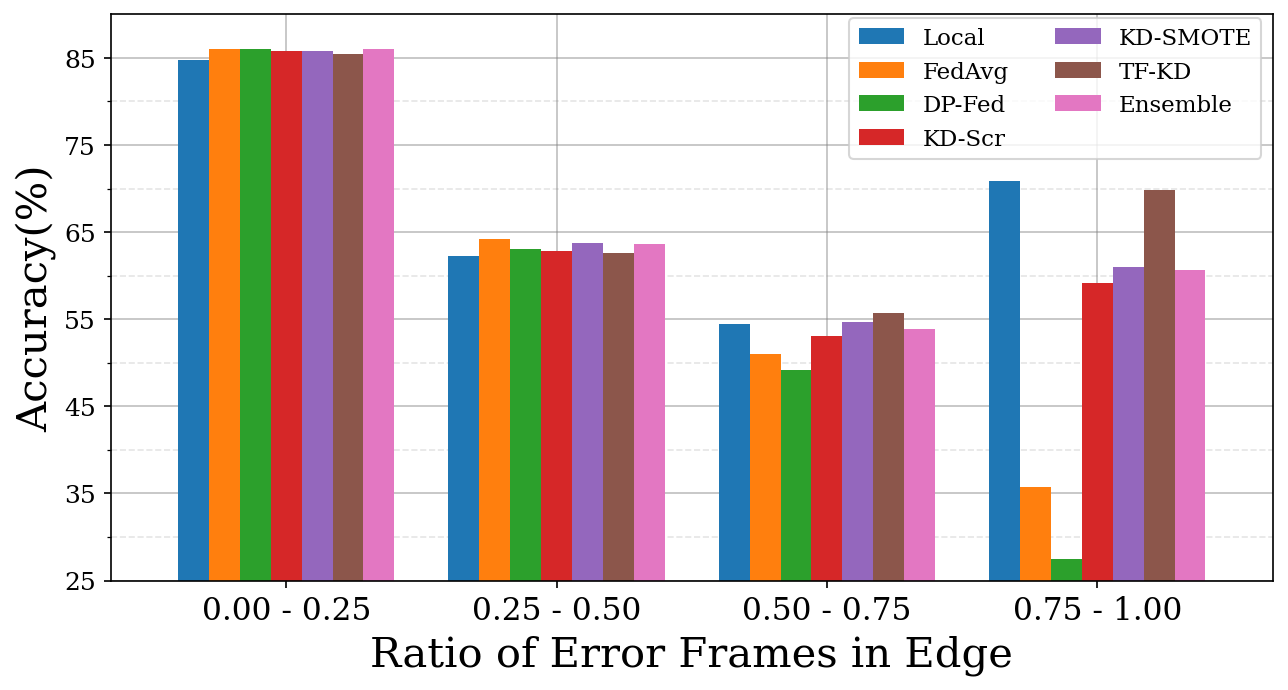}
    \caption{Accuracy over different node error ratios.}
    \label{fig:RatioAccr}
\end{figure}

\begin{table}
\centering
\captionsetup{justification=centering}
\caption{No. of nodes for Accuracy Delta ranges.}
\label{table1}
\begin{tabular}{| R{1.5cm} | C{.6cm} | C{.6cm} | C{.6cm} | C{.8cm} | C{.6cm} | C{.8cm} |}
\hline
   \textbf{$\Delta$ with Local Training}  &  \textbf{Fed-Avg} &  \textbf{DP-Fed} &  \textbf{KD-Scr} &  \textbf{KD-SM-OTE} &  \textbf{TF-KD} &  \textbf{Ense-mble} \\
   \hline
    < -25  &      21 &      26 &       4 &         4 &      4 &         3 \\
    \hline
    -15 to -25      &      22 &      26 &       6 &         4 &      3 &         5 \\
    \hline
    -15 to -5       &      36 &      41 &      46 &        41 &     18 &        26 \\
    \hline
    -5 to 5         &     400 &     390 &     455 &       450 &    490 &       478 \\
    \hline
    5 to 15         &      53 &      51 &      42 &        43 &     38 &        36 \\
    \hline
    15 to 25        &      10 &      14 &       4 &        12 &      3 &         7 \\
    \hline
    > 25 &      20 &      14 &       5 &         8 &      6 &         7 \\
    \hline
\end{tabular}
\end{table}

We then investigate the variability in performance across different edge nodes via considering local training as a baseline and show the results in Table \ref{table1}. Both scrimmages 4 and 5 are considered together. Here, a negative $\Delta$ value signifies better local training performance. We can observe that federated methods perform significantly worse ($\Delta < -25\%$ and $\Delta < -15\%$) for a large number of nodes, which indicates lack of reliability and larger variability in performance across different edge nodes, which is consistent with the intuition behind the bias issue hindering federated learning generalization performance \cite{agnostic_federated_learning}. Contrarily, \textbf{KD based methods, and specially TF-KD, maintain more uniform performance and distribute their gain more fairly across edge nodes}. 

In Fig. \ref{fig:RatioAccr}, we divide nodes into four groups based on frame error rate. \textbf{We note how federated learning becomes unreliable at nodes with high frame error rates}, as explicitly exploiting local data becomes crucial to learn the multitude of error causes. 
Finally, it is important to note how TF-KD outperforms other KD based methods at high frame error rates.

\section{concluding Remarks}
We investigated the potential of knowledge distillation for wireless edge learning, via an empirical comparison of frame error prediction performance with federated learning using a recently available DARPA SC2 dataset. The proposed framework further incorporates privacy preservation through relying on SMOTE generated, rather than actual, data at the cloud. Interestingly, the proposed framework, and specially the transfer learning based TF-KD method, overcome the weaknesses of popular federated learning methods, when applied to the considered spectrally congested wireless environments, and exhibit robustness against catastrophic failures and performance degradation with high frame error rates. This lays the ground for further studies on knowledge distillation based methods in wireless edge networks with highly variable dynamics affecting different nodes.


\ifCLASSOPTIONcaptionsoff
  \newpage
\fi

\bibliographystyle{IEEEtran} 

\bibliography{ssp2021}

\begin{thebibliography}{10}
\providecommand{\url}[1]{#1}
\csname url@samestyle\endcsname
\providecommand{\newblock}{\relax}
\providecommand{\bibinfo}[2]{#2}
\providecommand{\BIBentrySTDinterwordspacing}{\spaceskip=0pt\relax}
\providecommand{\BIBentryALTinterwordstretchfactor}{4}
\providecommand{\BIBentryALTinterwordspacing}{\spaceskip=\fontdimen2\font plus
\BIBentryALTinterwordstretchfactor\fontdimen3\font minus
  \fontdimen4\font\relax}
\providecommand{\BIBforeignlanguage}[2]{{%
\expandafter\ifx\csname l@#1\endcsname\relax
\typeout{** WARNING: IEEEtran.bst: No hyphenation pattern has been}%
\typeout{** loaded for the language `#1'. Using the pattern for}%
\typeout{** the default language instead.}%
\else
\language=\csname l@#1\endcsname
\fi
#2}}
\providecommand{\BIBdecl}{\relax}
\BIBdecl

\bibitem{mcmahan2017communication}
B.~McMahan, E.~Moore, D.~Ramage, S.~Hampson, and B.~A. y~Arcas,
  ``Communication-efficient learning of deep networks from decentralized
  data,'' in \emph{Artificial Intelligence and Statistics}.\hskip 1em plus
  0.5em minus 0.4em\relax PMLR, 2017, pp. 1273--1282.

\bibitem{melis2019exploiting}
L.~Melis, C.~Song, E.~De~Cristofaro, and V.~Shmatikov, ``Exploiting unintended
  feature leakage in collaborative learning,'' in \emph{2019 IEEE Symposium on
  Security and Privacy (SP)}.\hskip 1em plus 0.5em minus 0.4em\relax IEEE,
  2019, pp. 691--706.

\bibitem{shokri2017membership}
R.~Shokri, M.~Stronati, C.~Song, and V.~Shmatikov, ``Membership inference
  attacks against machine learning models,'' in \emph{2017 IEEE Symposium on
  Security and Privacy (SP)}.\hskip 1em plus 0.5em minus 0.4em\relax IEEE,
  2017, pp. 3--18.

\bibitem{mcmahan2017learning}
H.~B. McMahan, D.~Ramage, K.~Talwar, and L.~Zhang, ``Learning differentially
  private recurrent language models,'' \emph{arXiv preprint arXiv:1710.06963},
  2017.

\bibitem{agnostic_federated_learning}
\BIBentryALTinterwordspacing
M.~Mohri, G.~Sivek, and A.~Suresh, ``Agnostic federated learning,''
  \emph{CoRR}, vol. abs/1902.00146, 2019. [Online]. Available:
  \url{http://arxiv.org/abs/1902.00146}
\BIBentrySTDinterwordspacing

\bibitem{jameel2020deep}
A.~S. M.~M. Jameel, A.~P. Mohamed, X.~Zhang, and A.~E. Gamal, ``Deep learning
  for frame error prediction using a {DARPA Spectrum Collaboration Challenge
  (SC2)} dataset,'' \emph{arXiv preprint arXiv:2005.01446}, 2020.

\bibitem{abadi2016deep}
M.~Abadi, A.~Chu, I.~Goodfellow, H.~B. McMahan, I.~Mironov, K.~Talwar, and
  L.~Zhang, ``Deep learning with differential privacy,'' in \emph{Proceedings
  of the 2016 ACM SIGSAC conference on computer and communications security},
  2016, pp. 308--318.

\bibitem{chawla2002smote}
N.~V. Chawla, K.~W. Bowyer, L.~O. Hall, and W.~P. Kegelmeyer, ``Smote:
  synthetic minority over-sampling technique,'' \emph{Journal of artificial
  intelligence research}, vol.~16, pp. 321--357, 2002.

\bibitem{hinton2015distilling}
G.~Hinton, O.~Vinyals, and J.~Dean, ``Distilling the knowledge in a neural
  network,'' \emph{arXiv preprint arXiv:1503.02531}, 2015.

\bibitem{bamTech}
{DARPA SC2 Team BAM! Wireless}, ``Adaptive wireless networks for spectrally
  efficient communication: {DARPA SC2 Phase 2} technical paper,''
  \emph{available at
  https://web.ics.purdue.edu/~elgamala/BAMWireless/bamTech.pdf}, 2021.

\end{thebibliography}

\end{document}